# Directing Monolayer Tungsten Disulfide Photoluminescence using a Bent Plasmonic Nanowire on a Mirror Cavity

*Shailendra K. Chaubey*[*, †], *Sunny Tiwari*[*, †], *Ashutosh Shukla, Gokul M. A., Atikur Rahman, and G.V. Pavan Kumar*[*]

Department of Physics, Indian Institute of Science Education and Research, Pune-411008, India
*E-mail: shailendrakumar.chaubey@students.iiserpune.ac.in,
sunny.tiwari@students.iiserpune.ac.in, and pavan@iiserpune.ac.in

[†] Authors contributed equally.

**Abstract:** Designing directional optical antennas without compromising the field enhancement requires specially designed optical cavities. Herein, we report on the experimental observations of directional photoluminescence emission from a monolayer Tungsten Disulfide using a bent-plasmonic nanowire on a mirror cavity. The geometry provides field enhancement and directivity to photoluminescence by sandwiching the monolayer between an extended cavity formed by dropcasting bent silver nanowire and a gold mirror. We image the photoluminescence emission wavevectors by using Fourier plane imaging technique. The cavity out-couples the emission in a narrow range of wavevectors with a radial and azimuthal spreading of only 11.0° and 25.1°, respectively. Furthermore, we performed three dimensional finite difference time domain based numerical calculations to corroborate and understand the experimental results. We envisage that the results presented here will be readily harnessed for on-chip coupling applications and in designing inelastic optical antennas.

Controlling and manipulating spontaneous emission is a principal indispensable task in nanophotonics[1-5]. The emergence of cavity electrodynamics has provided an excellent framework to study and manipulate spontaneous emission[6-8]. Molecular fluorescence coupled to micro and nano-cavities has been widely studied from fundamental point of view as well as for application purposes[9-11]. Of late, studying spontaneous emission from emitters confined in plasmonic cavity have gained relevance. Specifically, plasmonic cavities formed using mirror substrates has gained special attention because of ease of preparation and very intense electric field inside the cavity formed between the structure and its image across the mirror substrate[10, 12-13].

To this end, various structures have been utilized for harnessing the field such as nanoparticle[12], nanocube[14], and nanowire[10, 15]. Out of these, nanowire not only provides enhancement but also directs the light from molecules sandwiched between the nanowire and mirror[10, 16]. Furthermore, bending the nanowire has shown to enhance the directionality of the secondary emission from molecules placed on the nanowire[17] or in the gap between the nanowire and mirror cavity[18]. The gap between the particle-image particle can be controlled by placing either molecular monolayer or 2d monolayers of finite thickness[9, 11, 17].

Recently two-dimensional transition metal dichalcogenides (TMDs) have emerged as new quantum emitters with high exciton binding energy, which lead to well pronounced optical transitions[19-22]. The direct bandgap of monolayers of TMDs makes them suitable for switching and optoelectronics devices[20, 23]. Well defined molecular orientation and dipole moment of a monolayer of TMDs make it a more suitable candidate to study quantum cavity electrodynamics[24-26]. TMDs show other unique optical properties such as strong spin-orbit interaction and valley polarization,[27-29] second harmonic generation,[30-32] chiral Raman[33]. Plasmonic and dielectric cavity coupled to TMDs has been utilized for strong coupling,[13, 34-36] photoluminescence enhancement,[37-39] tailoring the emission,[14] trion enhancement,[40] and chiral routing of Raman[41] and photoluminescence[42].

Motivated by this, we study a bent plasmonic nanowire on a mirror cavity with a monolayer of $WS_2$ sandwiched in the cavity. We show that the cavity can direct the $WS_2$ photoluminescence (PL) emission to a very narrow range of wavevectors when the emission is collected from the kink region, upon excitation of the nanowire end. Using Fourier plane imaging[43-46] we image the PL emission wavevectors, and corroborate and understand the experimental findings using three dimensional finite difference time domain based numerical calculations.

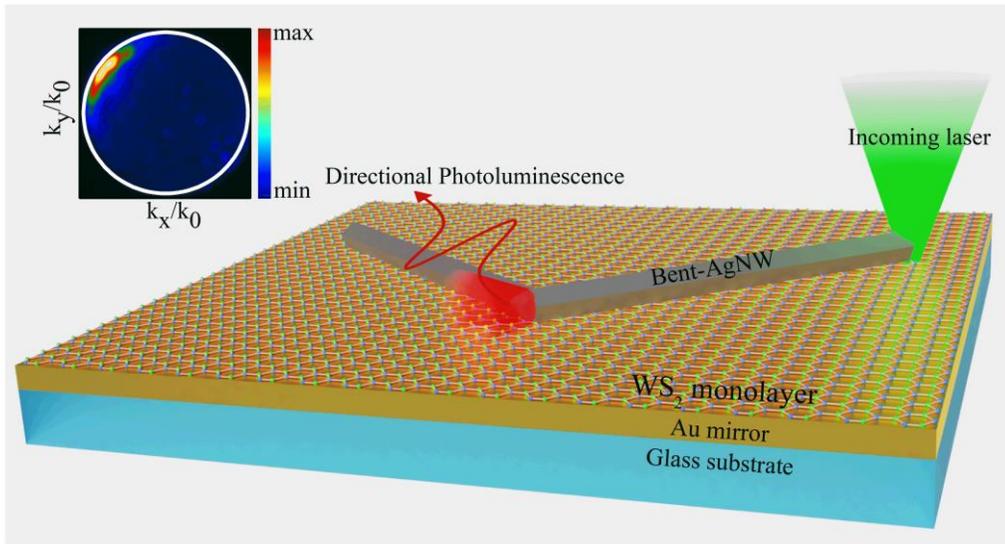

*Figure 1:* *Schematic representation of the experimental configuration. A monolayer of $WS_2$ is grown and transferred onto the top of a gold mirror. A bent-silver nanowire was dropcasted on a gold mirror thus forming a bent-nanowire on a mirror (B-NWoM) cavity with the monolayer in the cavity. One end of the nanowire was excited using a high numerical aperture objective lens and the emission from kink region was spatially filtered for spectroscopy and Fourier plane imaging. The emission from the kink is directional in nature and covers a very narrow range of angles.*

**Figure 1** shows the schematic of the experiment configuration. A monolayer of $WS_2$ was sandwiched between a gold mirror and a bent-silver nanowire thus confining the monolayer in an extended cavity. One end of the nanowire was excited using a focused laser source of wavelength 532 nm with polarization along the length of the nanowire. Propagating surface plasmon polaritons (SPPs) outcouple from the kink part of the nanowire, exciting the cavity enhanced monolayer photoluminescence (PL) of $WS_2$. The fundamental SPPs along the nanowire and incoming laser source at the nanowire end also excites the cavity assisted PL emission from the $WS_2$ which also get couple to the nanowire SPPs because of near-field coupling and out-couple from the kink part of the nanowire. The emission from the kink of B-NWoM cavity is collected using spatial filtering and is routed to the spectrometer for spectroscopy or EMCCD for Fourier plane imaging.

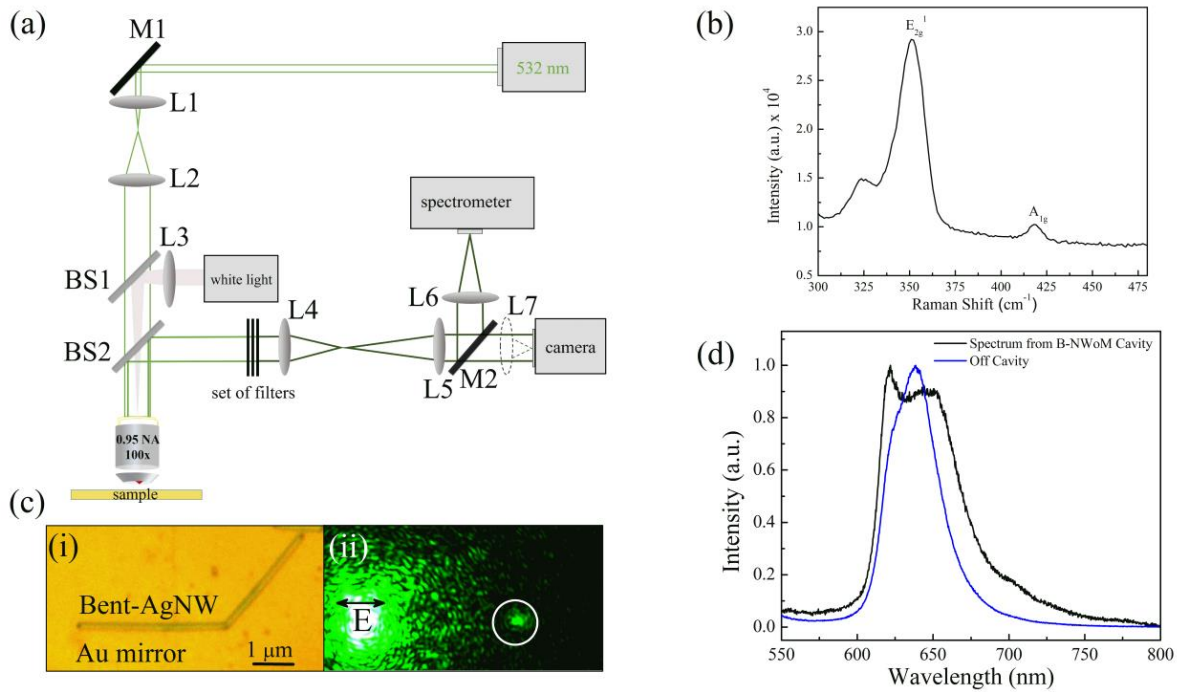

*Figure 2:* *Optical imaging and PL spectroscopy. (a) Schematic of the experimental setup used to probe the Fourier plane imaging of $WS_2$ PL from B-NWoM cavity. (b) Raman spectroscopy of the $WS_2$ layer confirming the monolayer. (c) Optical imaging of B-NWoM cavity. (i) Bright field image of a 350 nm thick bent nanowire placed on a gold mirror of thickness 160 nm with a monolayer of $WS_2$ between the gap. (ii) Elastic scattering image of the same bent nanowire when one end of the nanowire was excited with a 532 nm laser. (d) $WS_2$ PL spectrum collected from the kink part (shown as white circle in figure 2c (ii)) of the bent-nanowire upon excitation of one end of the bent nanowire (black curve) and spectrum collected only from the $WS_2$ monolayer placed on the gold mirror (blue curve).*

A detailed optical setup and imaging of the geometry is shown in **figure 2(a)**. The sample was excited using a high numerical aperture 100x, 0.95 NA objective lens. The backscattered light was collected using the same lens. The 532 nm laser light was expanded using a set of two lenses L1 and L2. M1 is a mirror. The polarization of the incoming laser was controlled by a λ/2 waveplate in the path. BS1 and BS2 were beam splitters to simultaneously excite the sample with laser and its visualization using white light. Lens L3 was used to loosely focus white light on the sample plane. A set of two edge filters and one notch filter were used to reject the elastically scattered light for PL spectroscopy and Fourier plane imaging. Lens L4 was used to recreate the real image for spatial filtering. Lens L5 was used to image the Fourier plane on the camera or CCD whereas lens L7 was used to flip lenses used to switch from real plane to Fourier plane. Mirror M2 and Lens L6 were used to route light towards the spectrometer and to focus the light onto the spectrometer, respectively.

Figure 2(b) shows the optical characterization of the monolayer $WS_2$. Monolayer $WS_2$ was grown using atmospheric pressure chemical vapor deposition (APCVD) at 300 nm $SiO_2$ coated silicon wafer following the procedures given in references[47-48]. Gold mirror was prepared by depositing 160 nm gold on a glass cover slip using thermal vapor deposition. Monolayer $WS_2$ was then transferred to the gold mirror using wet transfer method using polystyrene support film[49]. After the transfer, $WS_2$ layers were characterized by PL and Raman spectroscopy. Figure 2(b and d) shows the optical characterization of the monolayer $WS_2$. High PL intensity confirms the sample probed was a monolayer. In the Raman spectrum, the intensity of the $E_{2g}^1$ is much higher than $A_{1g}$ and the peaks $E_{2g}^1$ and $A_{1g}$ are positioned at 351 $cm^{-1}$ and 417 $cm^{-1}$, respectively, which matches with the reported Raman spectra for monolayer $WS_2$[50-51].

Silver nanowires (AgNWs) of average diameter 350 nm were synthesized using polyol process[52]. The nanowires were sonicated for 30 seconds to bend the nanowires[53]. Post bending the AgNWs were dropcasted on the $WS_2$ transferred gold mirror. This leads to the placement of the monolayer $WS_2$ in the cavity formed between the bent nanowire and mirror (B-NWoM cavity). The optical images of the monolayer $WS_2$ placed in the B-NWoM cavity is shown in figure 2(c). The bright field image shows a bent- silver nanowire of thickness ~350 nm is placed on a gold mirror. One end of the nanowire was excited using a focused laser source of wavelength 532 nm with polarization along the length of the nanowire to efficiently excite the propagating nanowire SPPs[54]. The light out couple from the kink portion of the nanowire end was spatially filtered for spectroscopy and Fourier plane imaging. Spectral signature collected from only the kink portion of the B-NWoM cavity shows a modulated spectrum as compared to the spectrum collected from the $WS_2$ monolayer placed only on the gold mirror (figure 2(d)). This modulation can be attributed to exciton to trion conversion cause by coupling of surface plasmon polaritons. Such exciton to trion conversion has been observed in metal-insulator-metal cavity by photoionization[16, 40].

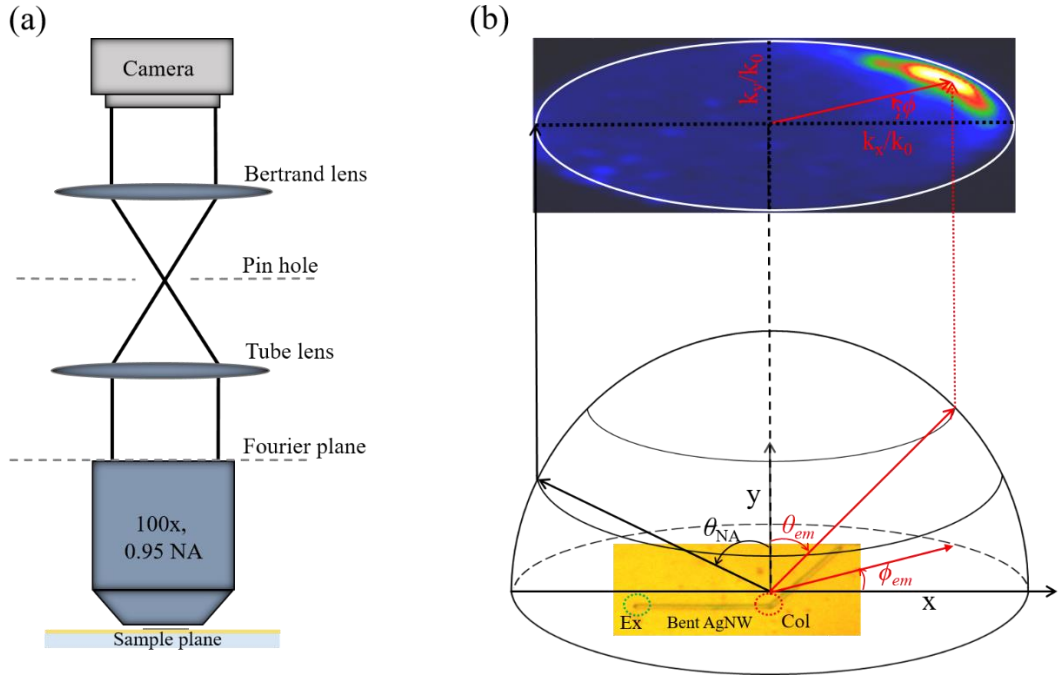

***Figure 3:*** *Schematic to visualize the Fourier plane imaging. (a) Experimental configuration to image the Fourier plane using objective lens. 4f configuration was used to reimage the Fourier plane from the back aperture of the objective lens to the camera. (b) The emission direction from the sample plane is marked as a point on the hemisphere. θ and ϕ are the radials and azimuthal coordinates in the Fourier plane. Schematic showing the direction emission in real plane (red curve) and Fourier plane is the two dimensional projection of the emission. Black curve shows the maximum angle corresponding to the numerical aperture of the objective lens.*

**Figure 3(a)** shows the schematic to visualize the directionality of the emission. The emission from the sample plane was collected using the objective lens, which forms the Fourier plane at the back aperture of the objective lens. To image the Fourier plane, it was recreated on the camera, using the combination of tube lens and Bertrand lens. Pin hole in the path was placed to selectively collect the emission from the kink part of the B-NWoM cavity using spatial filtering.

Figure 3(b) shows the schematic to understand the projection of the emission direction from the sample plane to the Fourier plane. Every point on the hemisphere represents a direction and can be written in terms of radial and azimuthal angles $\theta$ and $\phi$. The bent nanowire, placed in the x-y plane, emits the light in direction $\theta_{em}$ and $\phi_{em}$ as shown by red arrows. Experimental Fourier plane is the two dimensional projection of the emission in radial ($\theta$) and azimuthal angles ($\phi$). The directionality of the emission can be quantified by quantifying the spread in $\theta_{em}$ and $\phi_{em}$. Black curve shows the maximum angle ($\theta_{NA}$) corresponding to the numerical aperture of the objective lens.

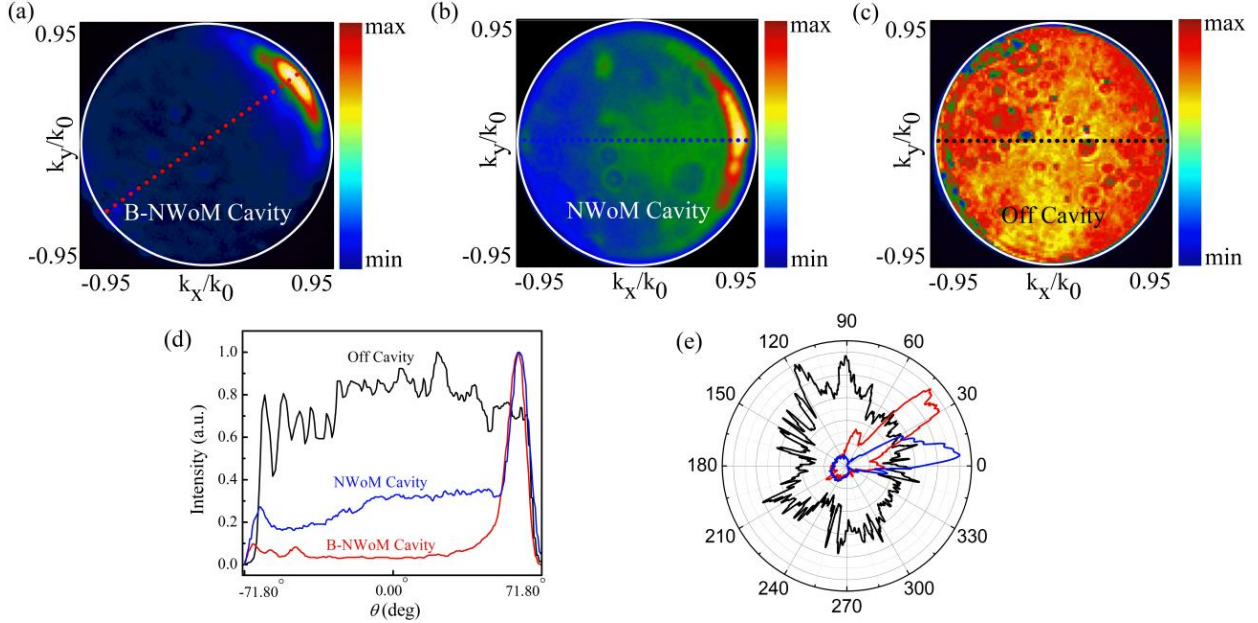

*Figure 4:* *Fourier plane imaging of* $WS_2$ *photoluminescence emission from B-NWoM cavity. (a) Fourier plane image of photoluminescence emission from the kink portion of B-NWoM cavity upon excitation of one end of nanowire after rejecting elastic scattering light. (b) Fourier plane image of photoluminescence emission from the distal end of a nanowire placed on a mirror cavity (B-NWoM cavity) upon excitation of one end of nanowire after rejecting elastic scattering light. (c) Fourier plane imaging of* $WS_2$ *photoluminescence emission from a monolayer of* $WS_2$ *placed on a gold mirror. The excitation and collection regions were the same. (d) Intensity cross-cuts along the red dotted line in (a), blue dotted line in (b), and black dotted line in (c). (d) Intensity profile of azimuthal angles (ϕ) for θ corresponding to maximum intensity in the Fourier plane images (a-b).*

To further study the $WS_2$ PL emission spectrum from the kink portion of B-NWoM cavity, we performed Fourier plane imaging on the $WS_2$ PL emission collected only from the kink portion. **Figure 4(a)** shows the Fourier plane image of the PL emission collected from the kink portion of the B-NWoM cavity using spatial filtering technique. The emission is confined to a narrow range of wavevectors and is directed towards the higher $k_x/k_0$ and $k_y/k_0$. On the other hand, the Fourier plane imaging performed on the PL emission from the distal end of the nanowire without the bending shows that the emission is directed towards higher angles and the majority of the emission happens along the nanowire length (figure 4 (b)). For comparison, the Fourier plane image of off-cavity emission from only the monolayer $WS_2$ on a gold mirror is shown in figure 4(c). The emission is isotropic in nature and covers approximately the full numerical aperture of the lens. The intensity cross-cut along the red, blue, and black dashed lines in figures 4(a-c) are shown in figure 4(d). The emission in the case of B-NWoM is very confined in radial angles with a full width at half maxima of 11.0°. The intensity profile of azimuthal angles (ϕ) for θ corresponding to maximum intensity in the Fourier plane images 4(a- c) are shown in figure 4(d). Along with the confinement in radial angles, the emission is also confined in terms of azimuthal angles in the case of B-NWoM cavity with a full width at half maxima of 25.1°. The nanowire on mirror (NWoM) cavity also facilitates the directional emission but with a slightly broader radial and azimuthal angular spreading, 11.4° and 31.7° respectively. Because of the finite reflection from the nanowire end the emission is also directed towards the $-k_y/k_0$ direction which results in the decrease in the forward to backward ratio of emission[54]. This is evident in the intensity cross-cut (blue curve in figure 4(d) where the emission is also finite near the maximum numerical aperture angles. On the other hand, the kink portion of the bent nanowire does not reflect the emission in the forward direction and out-couples the maximum of the mission in one direction only. This makes the B-NWoM cavity an excellent geometry for directing TMDs emission to a narrow range of angles.

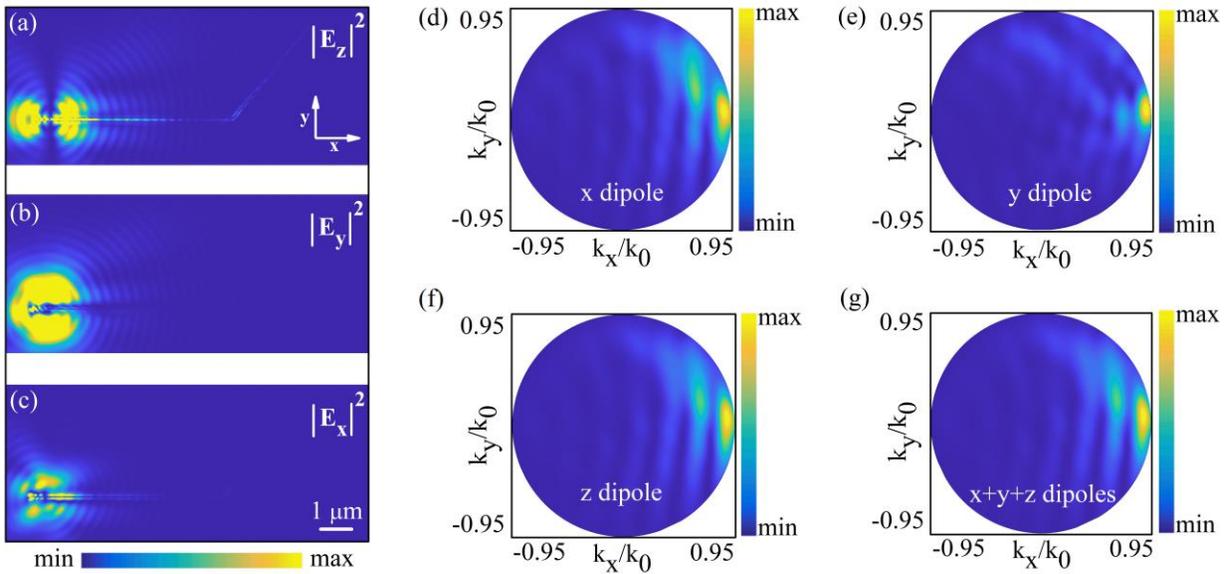

*Figure 5: FDTD based numerical calculations to understand the directional photoluminescence emission from B-NWoM cavity. (a-c) Component wise magnitude of near-field electric field in the B-NWoM cavity upon excitation of one end of nanowire with a focused Gaussian laser source of 532 nm. Calculated Fourier plane images of the emission from the kink portion of B-NWoM cavity, upon excitation of nanowire end by an oscillating dipole along x axis (d), y axis (e), z axis (f), and an incoherent sum of emission from all three dipoles (g). The dipoles are oscillating at a wavelength of 620 nm, which is near the maxima of monolayer $WS_2$ photoluminescence.*

To corroborate and understand the experimental results we performed FDTD based numerical calculations in Lumerical Software. The bent-nanowire was modelled by combining two cylindrical silver rods of length 6 µm. The radius of the cylinder was 125 nm and to make the kink portion, the radius was made to be 110 nm. The thickness of the gold mirror was set to be 100 nm. A 5 nm gap is set between the mirror and nanowire to account for the molecular monolayer thickness and the PVP coating on the nanowire[55]. Figure 5(a-c) shows the component-wise near-field electric field in the cavity upon illumination of one end of the nanowire using a high numerical aperture objective lens. The polarization of the incoming laser was set along the x axis, which is along the length of the nanowire for efficient excitation of the propagating nanowire SPPs. The $E_z$ field is much more enhanced in the cavity as compared to the $E_x$ and $E_y$ field. This enhances the emission from the $WS_2$ monolayer. To see how the molecular orientation affects the far-field emission wavevectors, we excite the cavity by placing a dipole oscillating at a wavelength of 620 nm at one of the ends of the bent-nanowire. A two dimensional monitor was placed near the kink part of the B-NWoM cavity and the near-field was projected to the far-field using the in-built feature in Lumerical software. Figure 5(d-f) shows the Fourier plane images when the end was excited by a dipole oscillating along the z, y, and x axis respectively. Figure 4(g) shows the Fourier plane image after the incoherent addition of the x, y, and z oriented dipole emission. In all the four cases the emission is direction in nature and the results match well with the experimentally obtained Fourier plane images. In the experimental configuration with a monolayer of $WS_2$, the dipolar contribution along the x and y axis are suppressed whereas the contribution of emission along the z axis get enhanced because of the presence of the mirror.

To conclude we show how a bent silver nanowire placed on a gold mirror can direct monolayer $WS_2$ PL which is otherwise isotropic in nature. The emission is confined to a narrow range of angles with a full width at half maxima in radial and azimuthal angles of 11.0° and 25.1° respectively. In addition to confining the emission wavevectors, the cavity broadens the PL spectrum by decoupling the exciton and trion contribution. Furthermore, using finite difference time domain based numerical calculations we understood the effect of dipole orientation of the two dimensional monolayer on the radiation pattern of PL emission. We believe that the results presented here will be extrapolated to route enhanced Raman signal and to design on-chip directional optical sources using various other class of two dimensional materials.

**Acknowledgement:**

Authors thank Diptabrata Paul for preparing the silver nanowires and in designing the schematic. This work was partially funded by DST Energy Science grant (SR/NM/TP-13/2016), Air Force Research Laboratory grant (FA2386-18-1-4118 R&D 18IOA118) and Swarnajayanti fellowship grant (DST/SJF/PSA02/2017-18) to G.V.P.K. AR acknowledges funding support from the Indo-French Centre for the Promotion of Advanced Research (CEFIPRA), project no. 6104-2.